\documentclass[manuscript]{aastex}

\slugcomment{\underline{Submitted to}: \apjl \hfill
\underline{Version}: \today}

\begin{document}

\title{The imprint of presupernova evolution on supernovae remnants}

\author{
Carles Badenes \email{badenes@ieec.fcr.es}
\and Eduardo Bravo \email{eduardo.bravo@upc.es}
}

\affil{Dept F\'\i sica i 
Enginyeria Nuclear, Universitat Polit\`ecnica de Catalunya, Av Diagonal 
647, 08028 Barcelona, and \\
Institut d'Estudis Espacials de Catalunya, Edifici 
Nexus,  Gran Capit\`a 2, 08034 Barcelona, Spain} 

\begin{abstract}
The evolution of type Ia supernova binary system progenitors is highly
uncertain. Several evolutionary models predict that the accretion of mass onto 
the white dwarf is accompanied by mass ejection from the binary in the form of a 
powerful wind, but very few observations have been made during the initial phase 
of 
formation of supernovae remnants, when the interaction of supernova ejecta with 
presupernova wind could be tested. Here we present hydrodynamical simulations of 
supernova ejecta interaction with an ambient medium modified by presupernova 
wind. The structure of the ambient medium when the supernova explodes 
is very sensitive to the details of wind history, and  
the evolution of the supernova remnant can be affected during several thousand 
years. We have found that the forward shock expansion parameter is a good tool 
for discriminating between several wind models.
The evolution of the supernova remnant in
the presence of an ambient medium modified by interaction with
pre-supernova wind cannot be described by a similarity solution. We also
rule out simple models based on a circumstellar medium that merges
smoothly with a uniform density ambient medium. 
\end{abstract}


\section{Introduction}

The evolution of type Ia supernova (SNIa) binary system progenitors is highly
uncertain. Although there is consensus that the exploding star is a white
dwarf made of carbon and oxygen, the nature of the companion star, the observational
counterpart of the binary, and the detailed evolutionary phases of the
system prior to SN explosion are currently the matter of strong debate (see,
for instance, Canal, M\'endez, \& Ruiz-Lapuente 2001, and references
therein). Those uncertainties raise doubts about the use of high-redshift
SNIa for cosmological purposes, especially as their calibration
is only based on the properties of local SNIa.
One way to discern among the various presupernova models is to look for their 
consequences on supernova remnant evolution. 

Models of supernovae remnants (SNR) originated by a SNIa usually
assume either a uniform
density interstellar medium (ISM) or a $\rho\propto r^{-2}$ circumstellar 
medium (CSM), 
produced by a constant mass loss rate wind \citep[for instance]{dc98}. This
assumption, together with
an appropriate choice of the SN ejecta profile, allows for the existence
of similarity
solutions that describe the early evolution of the SNR \citep{ch82,tm99}. Detailed 
models of pre-SNIa binary system evolution, however, point to a time varying 
mass loss rate \citep{hkn96,lan00}, which is expected to lead to ambient medium (AM) density 
profiles very 
different from a power law. 

Very few observations have been made during the initial phase of formation of 
supernovae remnants, when the interaction of supernova ejecta with presupernova 
wind could be tested. \citet{sp93} used ROSAT to search for X-ray
emission from the type Ia SN1992A 16 days after visual maximum, and derived an 
upper limit for the
presupernova mass-loss rate of less than a few times $10^{-6}
M_{\odot}{\rm yr}^{-1}$. \citet{cum96} looked for narrow H$\alpha$ in a
high-resolution spectrum of SN1994D, another SNIa, and derived an upper
limit for the mass loss rate of $\sim1.5\times10^{-5} M_{\odot}{\rm yr}^{-1}$. 
These observations have in common that they were made at a very early phase 
in the supernova
ejecta evolution,
implying that they only probed the CSM very close to the
progenitor.
The AM region up to a few parsecs can be probed by the evolution
of young SNR.

Here we present hydrodynamical simulations of supernova ejecta interaction with 
an AM modified by presupernova wind. We have worked out simple 
models of time dependent wind, devised to reproduce the history of wind 
given by pre-SNIa binary system evolutionary models. The main results 
of the simulations are presented and the compatibility of each wind model with 
known SNR from SNIa is discussed, based on the different outcomes of the 
calculations (mainly, expansion parameters, velocity of the ejecta at the reverse shock, 
and radius of the SNR).

\section{Hydrodynamical simulations of wind interaction with ambient medium
and SN ejecta}
 
\subsection{Wind models}

Current models of presupernova binary evolution describe wind ejection as a 
subproduct of matter accretion on the primary star, once it has become a white dwarf. 
In the last years, there has been a wide
interest on the possibility \citep{hkn96} that an optically thick wind 
stabilizes the
mass transfer in the binary system, thus opening a new channel for the
white dwarf to grow up to the Chandrasekhar mass and explode as a SNIa 
\citep{lv97,kv98,hkn99}. 
The time history of wind is complex, and depends on orbital parameters
as well as on the capability of the white dwarf to quietly accrete the
matter transferred from the secondary star. An important characteristic
of the wind history is whether the wind is active up to the time of SN
explosion or, on the contrary, the binary has a time extended phase of 
conservative
evolution prior to white dwarf ignition. Another important parameter is
the velocity of the wind. Here we explore the effects of
the wind on the SNR evolution, and compare with the results
obtained on the assumption that there is no significant presupernova
wind.

\begin{figure}
\epsscale{0.5}
\plotone{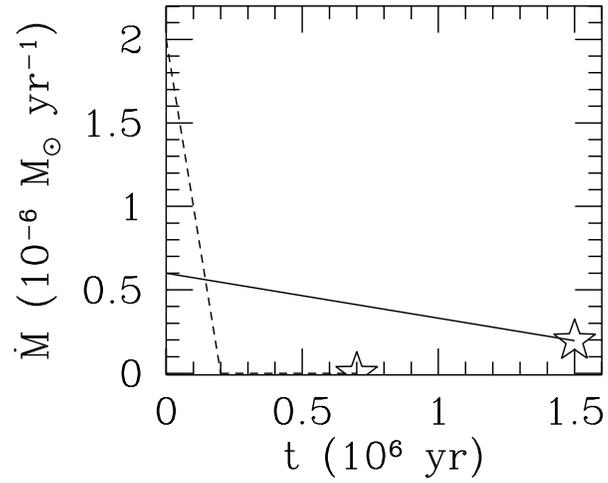}
\caption{Time evolution of wind mass loss rate for models A and B (dashed
line) and models C and D (solid line). The time at which the SN explodes
is identified by a $\star$. Models A and B are characterized by a phase
of conservative binary evolution prior to SN explosion}
\label{fig-1} 
\epsscale{1.0}
\end{figure}

We have worked out simple models of time dependent wind, devised to reproduce 
the gross features of 
the wind history predicted by binary system evolutionary models. In all our wind 
models the rate at which mass is lost by the binary system as a wind, 
$\dot{M}_{\rm w}$, has a linear time dependence:
$\dot{M}_{\rm w} = a - bt$, with $a$, $b$ two positive 
parameters, and the wind velocity, $v_{\rm w}$, remains constant with
time.
The wind can either be active until the moment of supernova 
explosion, as in models C (fast wind) and D (slow wind), 
or it can decrease to zero, leaving the system in a 
phase in which 
there is no wind ejection until the explosion ensues, as in models A
(fast wind) and B (slow wind). The parameters for the models are 
given in table~\ref{tbl-1} and figure~\ref{fig-1}. 

\subsection{Wind interaction}

We have studied the hydrodynamical processes induced by
the wind ejection: first, its interaction with
the AM and, second, the formation and evolution of the SNR.
The simulations were performed with a standard one dimensional
hydrodynamical code similar to that
described in \citet{tm99}, with the equations modified to include a
source of mass and momentum at the center, to simulate wind ejection. The
ISM density was set to $10^{-24}$~g cm$^{-3}$ in all cases. 

\begin{figure}
\epsscale{0.7}
\plotone{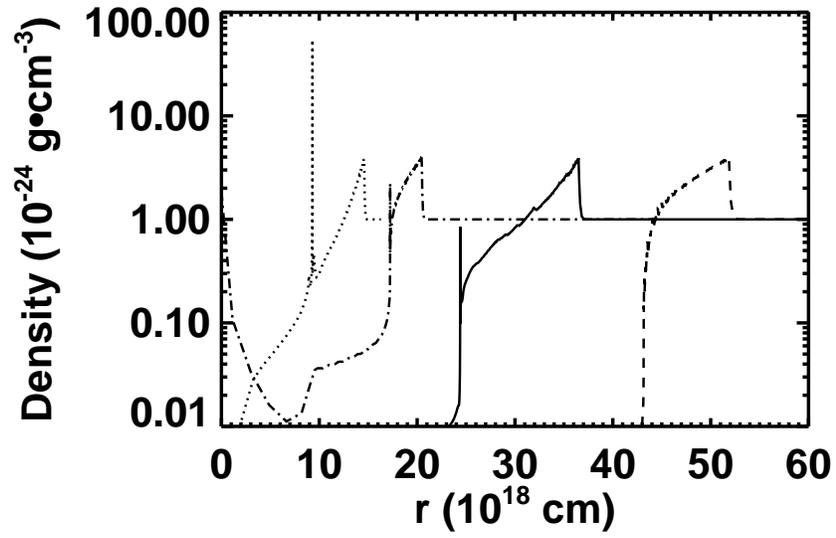}
\caption{Results of the hydrodynamical simulations: AM density profile at the 
time of SN explosion for the four wind models. Lines correspond to wind models A 
(solid), B (dotted), C (dashed), and D (dash-dotted)}
\label{fig-3} 
\epsscale{1.0}
\end{figure}

The resulting
AM density profiles at the time of SN explosion are shown in 
figure~\ref{fig-3}. 
The interaction of the wind ejected by the progenitor system and the
surrounding ISM follows a mechanism very similar to that at work in SNR:
high speed ejecta flow into a uniform, stationary medium and push it away
resulting in a doubly-shocked structure with a contact discontinuity
between the wind and the ISM. A forward shock (prominent in all the
profiles shown in figure~\ref{fig-3}) propagates into the ISM, heating,
compressing and accelerating it, while a reverse shock propagates inward,
heating, compressing and
decelerating the inner wind material. 
The kinetic energies involved,
however, are four to seven orders of magnitude lower than those
characteristic of SN explosions, and therefore the involved shock
velocities and temperatures are far below those typically found in SNR.
This kind of interaction between the wind and the ISM is not expected to
produce significant observational counterparts given the density and
temperature ranges involved.

The models with the highest wind velocity (A and C) produce a very large
bubble of low density AM around the SN. Model C presents the
largest bubble, because the time elapsed between the beginning of the wind and 
SN
explosion is the largest of the four models. One should expect
this feature to have a major impact on the evolution of the young SNR,
delaying the formation of the radiation emitting structures of the remnant.
Model D (low velocity wind) is the only one in which 
a power law structure
 can be identified close to the center. In this model, the reverse shock formed by the
interaction of the wind with the ISM lies at a radius of 3~pc, being the
only model in which
the reverse shock has not reached the center at the time of SN
explosion. 
Model B, characterized by a
low velocity wind with a
high mass loss rate, does not show a circumstellar region 
near to the
center because the wind stops blowing long before SN explosion.
This model presents the smoother density structure, broken
only by the presence of a contact discontinuity.

Once the SN explodes, and after the SNR characteristic doubly-shocked structure is formed, 
the
forward shock propagation, and therefore the whole dynamical evolution of
the SNR (shock velocities, expansion parameters, etc.) will depend
on the AM density gradient encountered. We have performed hydrodynamical
simulations of SNR evolution assuming an exponential model for the SN
ejecta density profile \citep{dc98}, with a total mass $1.4~M_{\odot}$
and a kinetic energy $K = 0.98\times10^{51}$~erg, and with the AM
structures obtained previously for the four wind models.

\begin{figure}
\epsscale{0.7}
\plotone{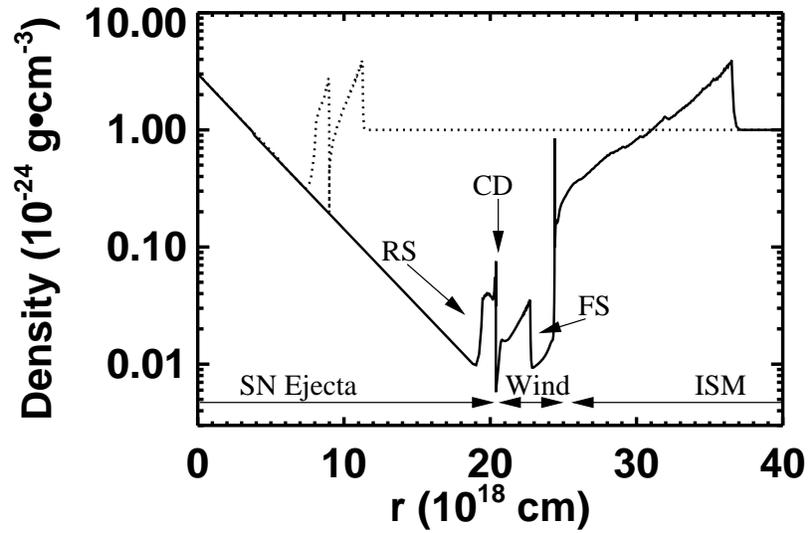}
\caption{Results of the hydrodynamical simulations of SNR
evolution at the age of Tycho's SNR. The SN ejecta was modelled by an 
exponential density
profile, and the wind by model A (solid line). Model EXPNW is also shown
(dotted line).
Labels RS, CD, and FS mark the positions of
the reverse shock, the contact discontinuity, and the forward
shock}
\label{fig-4} 
\epsscale{1.0}
\end{figure}

In figure~\ref{fig-4} we show 
the density profile at the age of Tycho's SNR for wind
model A (other wind models look similar, although details vary from case
to case, especially the time scale), together with the SNR obtained for the same 
exponential SN
ejecta evolving into a constant density AM (i.e. with no wind,
henceforth EXPNW). The
influence of the wind on the SNR is clearly seen: the location of
the doubly-shocked structure is farther away (and, for model A, clearly 
inconsistent
with Tycho's SNR) and its density is two orders of magnitude lower than in
the
EXPNW model. We stress that the structures that appear outside the region
delimited by the forward and the reverse shocks would remain undetectable
due to the low temperatures involved. 

The evolution of the SNR in presence of the AM modified by the wind is
far from being self-similar. For instance, once the forward shock runs into the 
contact
discontinuity between the wind and the ISM, it will experience a dramatic
change of behavior due to the strong positive density gradient. Another
change should be expected when the SNR forward shock runs into the wind
forward shock.
We emphasize that, depending on the wind model
adopted, these interactions can occur very early in the evolution of
the SNR.

\section{Discussion and conclusions}

We now proceed to analyze the implications of the computed hydrodynamical
evolutions on the observational properties of SNRs. We will concentrate
on a few relevant characteristics of SNRs: expansion
parameters, velocity at the reverse shock, and radius of the forward
shock.

\begin{figure}
\epsscale{1.1}
\plottwo{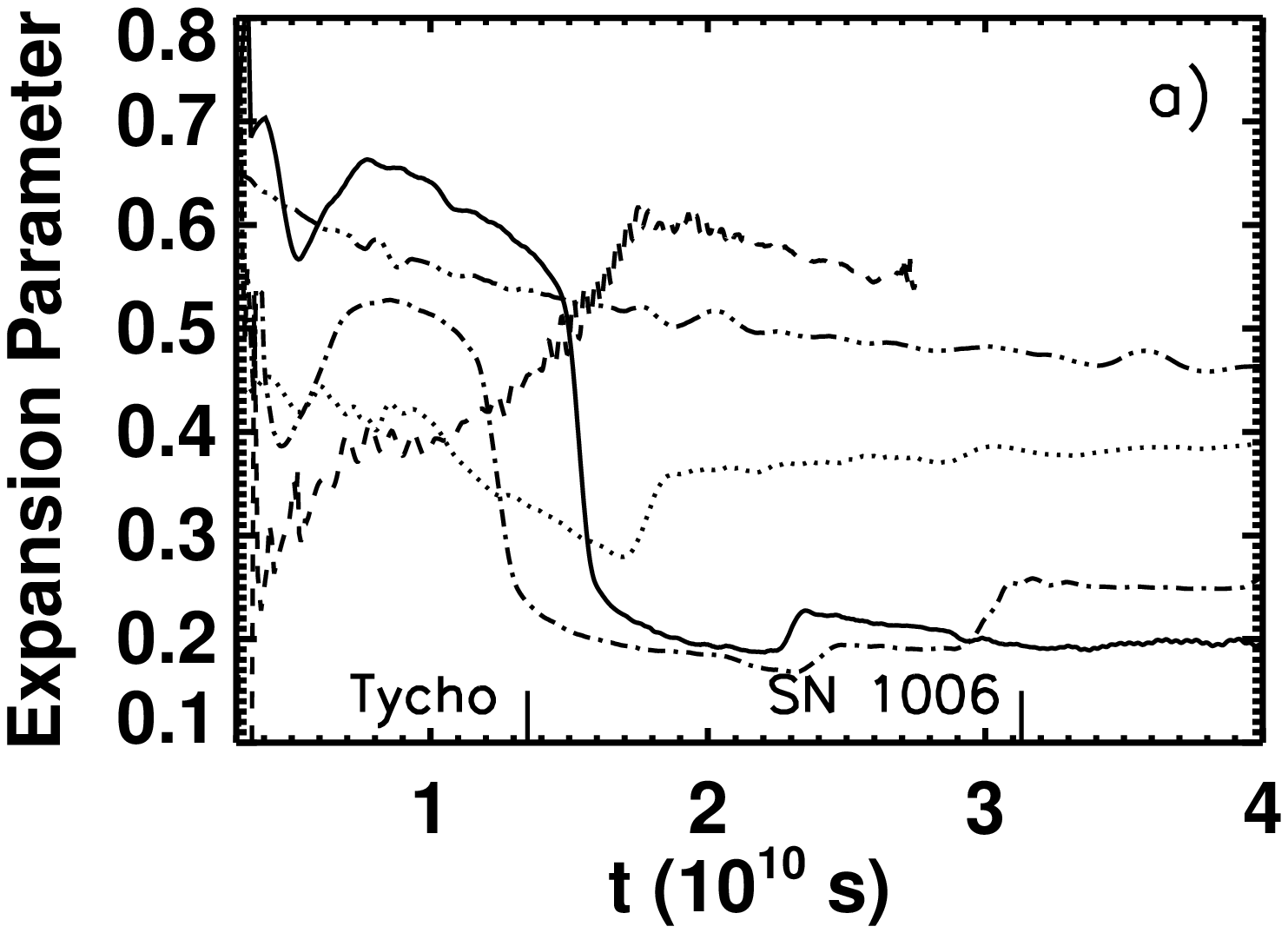}{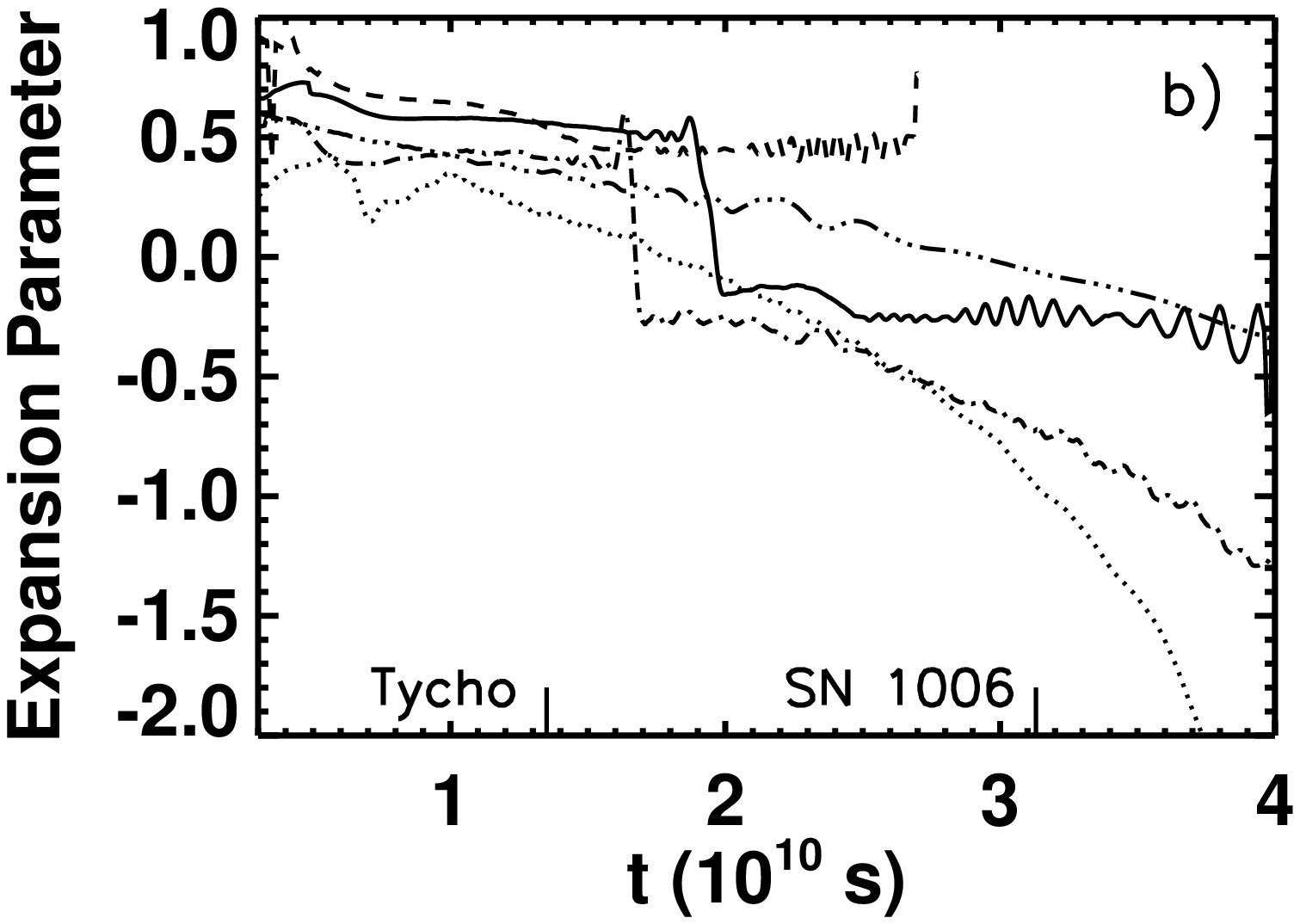}
\caption{Expansion parameter of the forward shock (a) and the reverse shock (b) 
as a function of time. The ages of Tycho's SNR and SNR 1006 are
marked above the horizontal axis. Lines correspond to wind models A 
(solid), B (dotted), C (dashed), D (dash-dotted), and model EXPNW
(dash-dot-dot-dotted)}
\label{fig-5}
\epsscale{1.0} 
\end{figure}

Figure~\ref{fig-5}a shows the temporal evolution of the expansion parameter of 
the SNR forward shock, which reflects the complexity of the AM
structure found by the SN ejecta. It is evident that the interaction of
the wind with the ISM influences the evolution of the SNR during several
thousands of years. At the end of the simulations ($\sim
4\times10^{10}$~s after SN explosion) only model EXPNW and
wind model B are close
to the Sedov phase (forward shock expansion parameter $\eta_{\rm
f}=0.4$). The evolution of wind models A and D is similar: they
experience a phase of rapid expansion, with values of $\eta_{\rm
f}$ comparable to the EXPNW model, until the forward shock reaches
the end of the low-density bubble, which causes the shock to slow down.
There follows a long phase in which $\eta_{\rm f}=0.2-0.3$ remains nearly
constant with slight speed ups.
The passage of the SNR forward shock through the wind-AM contact discontinuity 
produces a reflected component that rebounds at the ejecta-wind contact
discontinuity and overcomes the forward shock itself at 
$t \sim
2.3\times10^{10}$~s in models A and D, producing the first speed ups. 
The speed up at $t\sim
3\times10^{10}$~s (model D) is produced when the SNR forward shock
overcomes the wind forward shock 
(this feature is out of the figure time range for model A).
The speed up experienced by model B 
at $t \sim 1.8\times10^{10}$~s
is also due to the coalescence of the forward shock of the SNR with that due
to the interaction of the wind with the ISM. 
The simulation corresponding to model C was ended when the SNR forward
shock arrived to the end of the low-density bubble, due to 
the strong density contrast found there. 

Figure~\ref{fig-5}b shows the temporal evolution of the expansion parameter of 
the SNR reverse shock. As with the forward shock, in the wind models there are 
several secondary,
transmitted and reflected, shock waves that affect the evolution of the
SNR reverse shock, a feature that is not present in 
model EXPNW and that is less evident in wind model B, due to the
relative smoothness of its density profile. The sudden decrease of
the expansion parameter found in models A and D is caused by the arrival of
the secondary shock wave generated when the forward shock reaches the end
of the low-density bubble. Starting from the forward shock slow down
discussed earlier, this
reflected wave travels rapidly down the
negative density gradient of shocked wind, overcomes the ejecta-wind contact
discontinuity and, finally, coalesces with the SNR reverse
shock, giving it a push that drives 
$\eta_{\rm r}$ into negative values.

\begin{figure} 
\epsscale{1.1}
\plottwo{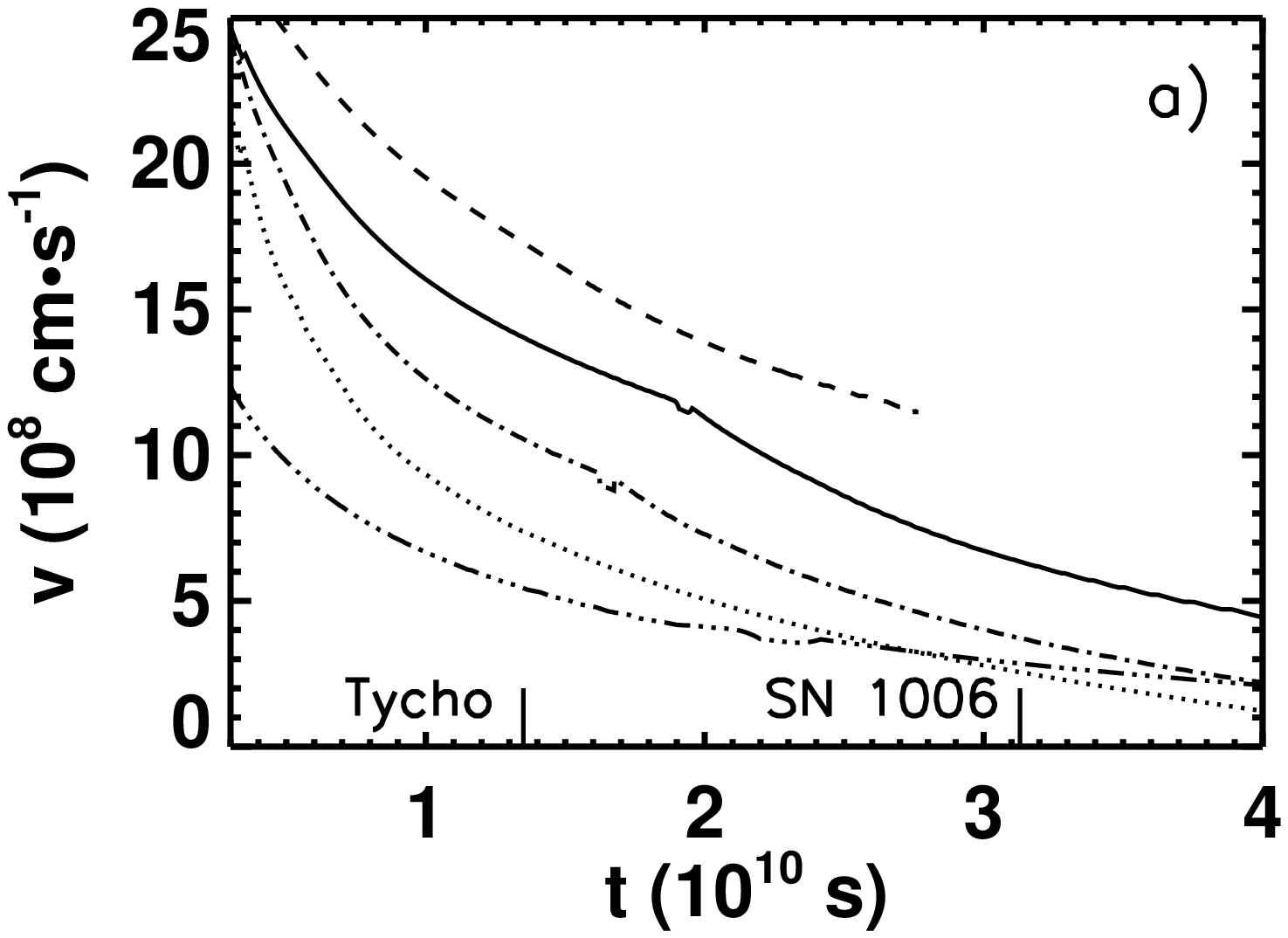}{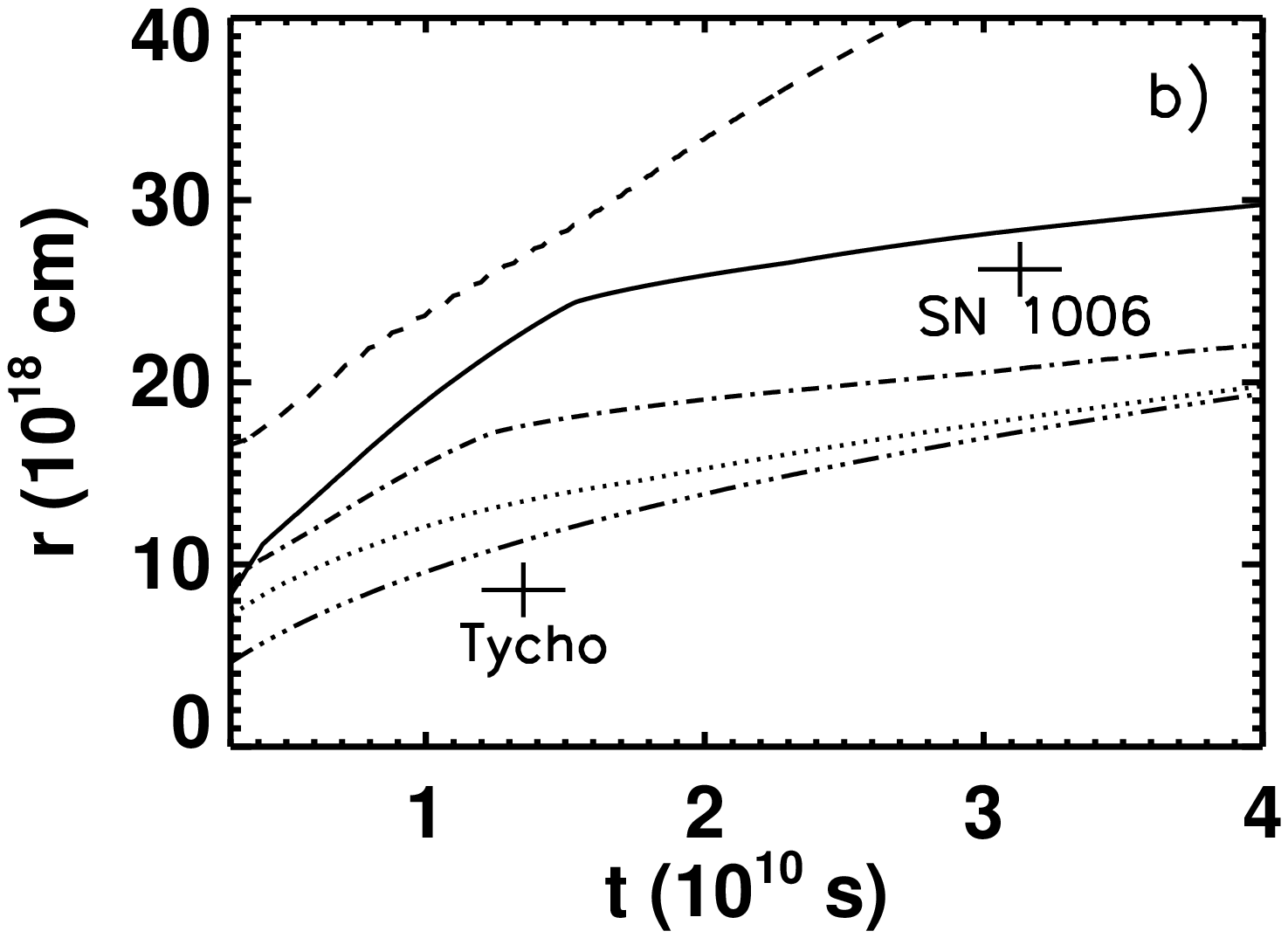}
\caption{Expansion velocity at the reverse shock (a) and radius
of the forward shock (b) as a function of time. The ages of Tycho's SNR 
and SNR 1006 are marked above the horizontal axis. Lines correspond to 
wind models A (solid), B (dotted), C (dashed), D (dash-dotted), and 
model EXPNW (dash-dot-dot-dotted)} 
\label{fig-6}  
\epsscale{1.0}
\end{figure}

The location of the reverse shock in lagrangian coordinate is better seen
in figure~\ref{fig-6}a, where the velocity of the SN
ejecta just ahead of the innermost shock is shown and compared to the EXPNW
model. The presence of low-density bubbles in all wind models
has two consequences. First, the build-up of the SNR shocks is delayed,
because the matter found by the expanding SN ejecta is so thin
that it takes a lot more time to influence the hydrodynamic evolution
of the ejecta.
This causes a delay in the penetration of the SNR reverse shock
into the SN ejecta (and, therefore a larger velocity compared to the EXPNW
model at the same epoch). Second, when the SNR reverse shock has already
been created, the density of SN ejecta that it encounters is lower in the wind
models, due to the longer expansion time. This results in a
more efficient inwards propagation of the reverse shock, which finally
converges with that of the EXPNW model.
In this figure, a change of the behavior
of the SNR reverse shock in models A and D can also be seen, which is due
to the coalescence of this shock and the one reflected in
the density barrier found at the end of the low-density bubble that we
have already mentioned.

Although we have made no attempt to fit the parameters of the wind models
to reproduce the properties of any particular SNR, the comparison of our
results with the well known SNR from Tycho's SN and SN1006 can be instructive. 
When comparing our models to real SNRs, it should be kept in mind that the details of the wind
and SNR evolution depend on the precise value chosen for the ISM density
at the beginning of the simulations (for instance, 
a value of the ISM density lower than we have chosen would result in a
larger low-density bubble and, thus, in a longer phase of high expansion
parameter, but also in larger velocities at the reverse shock and larger
radius of the SNR).

The average
expansion parameter of Tycho's SNR has been determined in radio and visible wavelengths
to be in the range $0.4-0.5$. \citet{h00} used {\sl ROSAT} X-ray
data fitted to analytic models of \citet{tm99} to obtain $\eta_{\rm f}=0.64$
and $\eta_{\rm r}=0.49$. While the reverse shock expansion parameter is
not particularly constraining with respect to the different wind models,
the high value found by \citet{h00} for the forward shock expansion
parameter can only be reproduced by wind models A and C (i.e., those with
the largest bubble) and, marginally, by the EXPNW model (we note
that Hughes found that his data could only be fitted by a model based on a
uniform density ejecta, but such a SN model is not realistic).
However, inspection of figure~\ref{fig-6}b shows that
Tycho's SNR radius is indeed too small when compared to models A and C. A 
decrease in the
initial value of the ISM density would lead to a larger expansion
parameter but also to a larger SNR radius. A spectral analysis of X-ray data
does
also pose constraints on the state of the SNR. \citet{dc98} suggested 
the existence of a CSM with $\rho\propto r^{-2}$ up to 0.67~pc from the
center merging smoothly into a uniform density AM, in order to explain the
spectroscopic features produced by Si, S and Fe. However, such a smooth 
structure
is in fact ruled out by our simulations of interaction of the wind with the
ISM.

With respect to SNR1006, the X-ray data point to a forward shock
expansion parameter $\eta_{\rm f}=0.5$, and a
velocity of the SN ejecta ahead of the reverse
shock of the order of $v = 4500-6300$~km
s$^{-1}$. Measurements in the visible range give a lower 
$\eta_{\rm f}$ of 0.33 \citep{lbv88}. The only model compatible with the
large value of $v$ is wind model A, which also gives a reasonable size for
SNR1006. On the other hand, wind model A is marginally compatible with
the expansion parameter found in the visible. One intriguing feature of
SNR1006 is the acceleration of cosmic rays up to the TeV energy range.
We suggest the possibility that the presence and interaction of several
secondary
shock waves, due to a low-density bubble produced by a wind could be
relevant to the explanation of the mechanism of acceleration of cosmic
rays in SNRs.

To conclude, we have shown that the wind ejected by the progenitor system
of a SNIa has a dramatic influence on the evolution of the corresponding
SNR, which is not correctly represented by the current assumption of a
CSM described by a simple power law and smoothly continued by a uniform
density AM. If current models of pre-SNIa
binary evolution are correct, the early evolution of their SNR cannot be
described by any kind of similarity solution. On the other hand, a few 
restrictions can also be put on
presupernova evolution. None of the wind models explored here is
compatible with the known properties of Tycho's SNR, whereas wind model
A (characterized by a high velocity wind lasting for about $2\times
10^5$~yr, followed by a phase of conservative evolution) gives results consistent with several features of SNR1006. It is
clear that a more complete exploration of the space of parameters
describing pre-supernova wind is of high interest. 

\acknowledgements 
This work has been supported by the MCYT grants ESP98-1348 and AYA2000-1785, and 
by the DGES grant PB98-1183-C03-02. CB is very indebted for a CIRIT 
grant.

\clearpage

\begin{deluxetable}{rrrrrrr}
\tabletypesize{\small}
\tablewidth{0pt}
\tablecaption{Parameters of wind models A, B, C, D \label{tbl-1}}
\tablehead{
\colhead{$a$} &
\colhead{$b$} & 
\colhead{$t_{\rm end}$} & 
\colhead{$t_{\rm SN}$} & 
\colhead{$v_{\rm w}$} &
\colhead{$M_{\rm w}$} & 
\colhead{$K_{51}$}  
} 
\startdata
$2$ & $10$ & $0.2$ & $0.7$ & $200$ & $0.2$ & $8\, 10^{-5}$ \\ 
$2$ & $10$ & $0.2$ & $0.7$ & $20$ & $0.2$ & $8\, 10^{-7}$ \\
$0.6$ & $0.27$ & $1.5$ & $1.5$ & $200$ & $0.6$ & $2.4\, 10^{-4}$ \\
$0.6$ & $0.27$ & $1.5$ & $1.5$ & $20$ & $0.6$ & $2.4\, 10^{-6}$ \\
\enddata
\tablecomments{Units of $a$ and $b$ are $10^{-6} M_{\odot}{\rm yr}^{-1}$ and 
$10^{-12} M_{\odot}{\rm yr}^{-2}$, respectively, $t_{\rm end}$ is the duration 
of 
the wind phase and $t_{\rm SN}$ is the time of the SN explosion, both in Myr, 
$v_{\rm w}$ is the wind velocity, in km/s, $M_{\rm w}$ is the total mass ejected 
in the wind, in $M_{\odot}$, $K_{51}$ is the total kinetic energy of the wind, 
in $10^{51}$~erg}
\end{deluxetable}

\clearpage

\clearpage

\clearpage

\clearpage


\begin{thebibliography}

\bibitem[Canal, M\'endez \& Ruiz-Lapuente(2001)]{kmr01} Canal, R.,
M\'endez, J., \& Ruiz-Lapuente, P. 2001, ApJ, 550, L53

\bibitem[Chevalier(1982)]{ch82} Chevalier, R. A. 1982, ApJ, 258, 790

\bibitem[Cumming et al.(1996)]{cum96} Cumming, R. J., Lundqvist, P.,
Smith, L. J., Pettini, M., \& King, D. L. 1996, \mnras, 283, 1355

\bibitem[Dwarkadas \& Chevalier(1998)]{dc98} Dwarkadas, V. V., \& Chevalier, R. 
A. 1998, ApJ, 497, 807 

\bibitem[Hachisu, Kato, \& Nomoto(1996)]{hkn96} Hachisu, I., Kato, M.,
\& Nomoto, K. 1996, ApJ, 470, L97

\bibitem[Hachisu, Kato, \& Nomoto(1999)]{hkn99} Hachisu, I., Kato, M.,
\& Nomoto, K. 1999, ApJ, 522, 487

\bibitem[Hughes(2000)]{h00} Hughes, J. P. 2000, ApJ, 545, L53

\bibitem[King \& van Teeseling(1998)]{kv98} King, A. R., \& van
Teeseling, A. 1998, \aap, 338, 965

\bibitem[Li \& van den Heuvel(1997)]{lv97} Li, X. D., \& van den Heuvel,
E. P. J. 1997, \aap, 322, L9

\bibitem[Langer et al.(2000)]{lan00} Langer, N., Deutschmann, A.,
Wellstein, S., \& H\"offlich, P. 2000, \aap, 362, 1046

\bibitem[Long, Blair \& van den Bergh(1988)]{lbv88} Long, K. S., Blair,
W. P., \& van den Bergh, S. 1988, ApJ, 333, 749

\bibitem[Schlegel \& Petre(1993)]{sp93} Schlegel, E. M., \& Petre, R.
1993, ApJ, 412, L29

\bibitem[Truelove \& McKee(1999)]{tm99} Truelove, J. K., \& McKee, C. F.
1999, \apjs, 120, 299

\end{thebibliography}
\end{document}